\documentclass[10pt]{article}
\usepackage{amsmath}
\usepackage{amsfonts}

\usepackage{amsmath,amssymb}
\usepackage{graphicx}
\usepackage{color}
\usepackage{url}

\newtheorem{theorem}{Theorem}
\newtheorem{lemma}{Lemma}
\newtheorem{corollary}{Corollary}

\newtheorem{proposition}{Proposition}
\newtheorem{remark}{Remark}
\newtheorem{conjecture}{Conjecture}

\newcommand{\F}{\ensuremath{\mathbb F}}

\newcommand{\done}{\hfill $\Box$ }

\newcommand{\Tr}{{{\rm Tr}}}
\newcommand{\N}{{{\rm N}}}

\newcommand{\ls}[1]
    {\dimen0=\fontdimen6\the\font\lineskip=#1\dimen0
     \advance\lineskip.5\fontdimen5\the\font
     \advance\lineskip-\dimen0
     \lineskiplimit=0.9\lineskip
     \baselineskip=\lineskip
     \advance\baselineskip\dimen0
     \normallineskip\lineskip\normallineskiplimit\lineskiplimit
     \normalbaselineskip\baselineskip
     \ignorespaces}

\begin{document}

\bibliographystyle{abbrv}

\title{Several Classes of  Permutation Trinomials over $\F_{5^n}$ From \\Niho Exponents }

\author{Gaofei Wu
\thanks{G. Wu is   with the State Key Laboratory of Integrated Service Networks,
Xidian University, Xi'an, 710071, China.
 Email: wugf@nipc.org.cn},
Nian Li
\thanks{N. Li is  with the  Department of Informatics, University of Bergen,
 N-5020 Bergen, Norway.
 Email:
nianli.2010@gmail.com.}
}
\date{}
\maketitle
\ls{1.5}

\thispagestyle{plain} \setcounter{page}{1}

\begin{abstract}
The construction of  permutation trinomials over finite fields attracts people's
interest recently due to their simple form and some additional properties.
Motivated by some results on the construction of permutation trinomials with Niho exponents,
by constructing some new fractional polynomials that  permute the set of the $(q+1)$-th roots of  unity in
$\F_{q^2}$,
we present several classes of permutation trinomials with Niho exponents over $\F_{q^2}$,
    where $q=5^k$.

{\bf Index Terms } Finite fields, Permutation polynomials, Trinomials, Niho exponents.

\end{abstract}

\ls{1.5}

\section{Introduction}


Let $p$ be a prime and $n$ be a positive integer.
Let $\F_q$  be a finite field of $q=p^n$ elements. We denote
$\F_q^* $ the multiplication group of $\F_{q}$.
A polynomial $f\in \F_{q}[x]$ is called a permutation polynomial  if
the associated polynomial mapping
$f:c\mapsto f(c)$ from $\F_{q}$ to itself is a permutation of
$\F_{q}$ \cite{Lidl97}.
Permutation polynomials over finite fields  have important applications
in cryptography \cite{Dobbertin99,Qu13}, coding theory \cite{Laigle07}, and combinatorial design theory \cite{Ding06}. Thus, it
is important to find new permutation polynomials, both theoretically and in practice.

Permutation polynomials with few terms especially binomials and trinomials
  attracts people's interest recently due
to their simple algebraic form and additional extraordinary properties.
Recently, there are  lots of constructions of permutation trinomials
over finte fields, please refer to \cite{Ding15,Gupta16,Hou15,Hou14,Hou152,Likangquan17,Likangquan16,Li161,Li162,Li163,Tu13,Zha17}.

In this paper, we focus on the  construction of permutation trinomials over $\F_{5^n}$
 from Niho exponents
with the form of
\begin{eqnarray}\label{polyform}
f(x)=x+\lambda_1x^{c_1(5^k-1)+1}+\lambda_2x^{c_2(5^k-1)+1},
\end{eqnarray}
where $n=2k$,  $ 1\leq s,t \leq 5^k$, and $\lambda_1,\lambda_2\in \{1,-1\}$.
  A positive integer $d$ (always understood modulo $p^n-1$) is a
Niho exponent if $d\equiv p^j \,({\rm mod}\, p^{\frac{n}{2}}-1)$ for
some $j<n$ \cite{Niho72}.
By finding some new classes of fractional polynomials that permute the set of
the $(5^k+1)$-th roots of unity in $\F_{5^{2k}}$,
we construct several classes of permutation trinomials of the  form \eqref{polyform}
 over $\F_{5^{2k}}$.

 The remainder of this paper is organized as follows. Section \ref{pre}
 introduces some notations and background. In Section \ref{mainres}, we present
 some classes of permutation trinomials over $\F_{5^n}$  by finding some fractional polynomials that permute  the set of
the $(5^k+1)$-th roots of unity in $\F_{5^{2k}}$.
 Some concluding remarks are given in  Section \ref{conclu}.

\section{Preliminaries}\label{pre}

Let $p$ be  prime, $n,\, k$ be two integers such that
$k~|~n$. The trace function from
$\F_{p^n} $ onto $\F_{p^k}$ is defined by
$$
\Tr_k^n(x)=\sum_{i=0}^{n/k-1}x^{p^{ik}}, \, x\in \F_{p^n}.
$$

\begin{lemma}\label{tracelemma}
Denote  by  $\Tr(x)$ and $\N(x)$ the trace function and norm function from $\F_{q^2}$ to $\F_{q}$,
where $q=5^k$.
Then
$$\Tr(x^2)=\Tr^2(x)-2\N(x),  ~\Tr(x^3)=\Tr^3(x)+2\N(x)\Tr(x),$$
 $$\Tr(x^4)=\Tr^4(x)+\N(x)\Tr^2(x)+2\N^2(x),$$
$$\Tr(x^8)=\Tr^8(x)+2\N(x)\Tr^6(x)-\N^3(x)\Tr^2(x)+2\N^4(x),$$ and
 $$\Tr(x^9)=\Tr^9(x)+\N(x)\Tr^7(x)+4\N^4(x)\Tr(x)+2\N^2(x)\Tr^5(x).$$
\end{lemma}
Lemma \ref{tracelemma} can be proved by direct computation and the proof is
omitted here.

The following lemma, which gives a powerful criterion of the polynomials of the form
 $x^lg(x^{\frac{p^n-1}{s}})$ to be permutation polynomials,  will be used to prove our  main results.

\begin{lemma}\cite{Park01,Zieve09}\label{lemkey}
Let $p$ be a prime.  Let $l,\, n $ and $s$ be positive integers such that $s~|~(p^n-1).$
Let $g(x)\in \F_{p^n}[x].$
Then $f(x)=x^lg(x^{\frac{p^n-1}{s}})$ is a permutation polynomial  over $\F_{p^n}$ if and only if
\begin{enumerate}
  \item [1)] $\gcd(l,\frac{p^n-1}{s})=1$, and
  \item [2)] $x^lg(x)^{\frac{p^n-1}{s}}$ permutes  $\mu_s,$
\end{enumerate}
 where $\mu_s$ is the set of $s$-th roots of unity in $\F_{p^n}.$
\end{lemma}

\section{New classes of permutation trinomials with Niho exponents} \label{mainres}

In this section, we present several classes of fractional polynomials that  permute   the set of
the $(q+1)$-th roots of unity in $\F_{q^2}$, then by Lemma \ref{lemkey}, we construct several classes of
permutation trinomials of the form \eqref{polyform}.

For simplicity,  throughout this section,   we  denote
$\Omega_{+}=\{x^2~|~x\in\mu_{q+1}\}$ and $\Omega_{-}=\{-x^2~|~x\in\mu_{q+1}\}$.

\begin{lemma}\label{q+3div4lem}
  Let $q=5^k$, where $k$ is a positive integer. Let  $g_1(x)=-x^{\frac{q+1}{2}}(\frac{x^s-2}{x^s+2})^{2}$,
   where $s=\frac{q+3}{4}.$ Then $g_1(x)$ permutes $\mu_{q+1}$.
\end{lemma}
{\em Proof:}
Note that  $ x^{\frac{q+1}{2}}=\pm 1$.
In the following we show that if
$x^{\frac{q+1}{2}}=1$, i.e., $x\in \Omega_{+}$,  then $-x^{\frac{q+1}{2}}(\frac{x^s-2}{x^s+2})^{2}
=-(\frac{x^s-2}{x^s+2})^{2}$ permutes $\Omega_{+}$. Similarly, if
$x^{\frac{q+1}{2}}=-1$, then $-x^{\frac{q+1}{2}}(\frac{x^s-2}{x^s+2})^{2}
=(\frac{x^s-2}{x^s+2})^{2}$ permutes $\Omega_{-}$. Then the
conclusion follows from $\Omega_{+} \bigcup \Omega_{-}=  \mu_{q+1} $
and $\Omega_{+} \bigcap \Omega_{-}= \emptyset.$

By  $x^{\frac{q+1}{2}}=1$, i.e., $x\in  \Omega_{+}$,  we have
$$(-(\frac{x^s-2}{x^s+2})^{2})^{\frac{q+1}{2}}=-(\frac{x^s-2}{x^s+2})^{q+1}
=-(\frac{x^{-s}-2}{x^{-s}+2})(\frac{x^s-2}{x^s+2})=
-(\frac{2+x^s}{2-x^s})(\frac{x^s-2}{x^s+2})=1,
$$
which shows that if $x\in \Omega_{+}, $ then  $ -(\frac{x^s-2}{x^s+2})^{2}\in \Omega_{+}. $
Suppose that there exist two different  elements $x_1=y_1^2,x_2=y_2^2\in \Omega_{+}$ such that
$-(\frac{x_1^s-2}{x_1^s+2})^{2}=-(\frac{x_2^s-2}{x_2^s+2})^{2}$.
Then
$\frac{x_1^s-2}{x_1^s+2}= \pm \frac{x_2^s-2}{x_2^s+2}$.
\begin{enumerate}
  \item [1)] If $\frac{x_1^s-2}{x_1^s+2}=  \frac{x_2^s-2}{x_2^s+2}$, then
$1+\frac{1}{x_1^s+2}=1+\frac{1}{x_2^s+2},$ i.e., $x_1^s=x_2^s$. Since
 $x_1^s=y_1^{2s} =y_1^{\frac{q+1}{2}+1} $ and  $x_2^s=y_2^{2s}=y_2^{\frac{q+1}{2}+1}$,
we have  $y_1=\pm y_2$ due to $y_1^{\frac{q+1}{2}}=\pm 1$ and $y_2^{\frac{q+1}{2}+1}=\pm 1$. Therefore,
$ x_1=x_2$.
  \item [2)] If  $\frac{x_1^s-2}{x_1^s+2}= - \frac{x_2^s-2}{x_2^s+2}$,
then $2+\frac{1}{2+x_1^s}+\frac{1}{2+x_2^s}=0$, i.e.,
\begin{eqnarray}\label{lem1eq}
2(2+x_1^s)(2+x_2^s)+2+x_1^s+2+x_2^s=0.
\end{eqnarray}
By \eqref{lem1eq}, we get $x_1^sx_2^s=-1$. Note that
$x_1^s=y_1^{2s}=y_1^{\frac{q+1}{2}+1}=\pm y_1$ and $x_2^s=y_2^{2s}=y_2^{\frac{q+1}{2}+1}=\pm y_2$,  we obtain
$y_1y_2=\pm 1$, thus  $x_1x_2=y_1^2y_2^2=1$, a contradiction with $(x_1x_2)^s=-1$.

%
\end{enumerate}
Based on the above analysis,    $-(\frac{x^s-2}{x^s+2})^{2}$ permutes $\Omega_{+}$.
 \done

\begin{theorem}\label{th1}
Let $q=5^k$,   $s=\frac{q+3}{4}, $ and $ t=2s=\frac{q+3}{2} $. Then
  $x+x^{s(q-1)+1}-x^{t(q-1)+1} $ permutes $\F_{q^2}$.
\end{theorem}
{\em Proof:}
According to Lemma \ref{lemkey},
it is sufficient to show that
$x(1+x^s-x^{t})^{q-1}$ permutes $\mu_{q+1}$.
First we prove that $1+x^s-x^{t} \neq 0 $ for any $x\in \mu_{q+1}$.
Otherwise, $ 1+x^s-x^{t} =-(x^s+2)^2=0$, i.e., $x^s=-2$. However,
$(x^s)^{q+1}=(-2)^{q+1}=-1$, a contradiction with $x\in \mu_{q+1}$.

Now suppose that  $1+x^s-x^{t} \neq 0 $ for any $x\in \mu_{q+1}$. Thus,
\begin{eqnarray*}
x(1+x^s-x^{t})^{q-1} &=&x\frac{(1+x^s-x^{2s})^q}{1+x^s-x^{2s}}=\frac{x(1+x^{-s}-x^{-2s})}{1+x^s-x^{2s}} \\
&=&-x^{1-2s}(\frac{x^s-2}{x^s+2})^{2}=-x^{-\frac{q+1}{2}}(\frac{x^s-2}{x^s+2})^{2}.
\end{eqnarray*}
Then the conclusion follows from Lemma \ref{q+3div4lem} and
$x^{-\frac{q+1}{2}}=x^{\frac{q+1}{2}}$.
\done

\begin{corollary}
Let $q=5^k$.  Then $f(x)=x^q+x^{(\frac{3}{4}(q-1)+2)(q-1)+1}-x^{\frac{q+1}{2}(q-1)+1}$
is a permutation trinomial over $\F_{q^2}$.
\end{corollary}
{\em Proof:} According to Lemma \ref{lemkey},
it is sufficient to prove that $g_2(x)=x(x+x^{\frac{3}{4}(q-1)+2}-x^{\frac{q+1}{2}})^{q-1}$
permutes $\mu_{q+1}$.
Note that $x^{-\frac{q+1}{2}}=x^{\frac{q+1}{2}}$, we have
\begin{eqnarray*}
g_2(x)&=&x(x+x^{\frac{3}{4}(q-1)+2}-x^{\frac{q+1}{2}})^{q-1}=x\frac{x^{-1}+x^{-\frac{3}{4}(q-1)-2}-x^{-\frac{q+1}{2}}}{x+x^{\frac{3}{4}(q-1)+2}-x^{\frac{q+1}{2}}}\nonumber \\
&=&\frac{1+x^{q+1}\cdot x^{-\frac{3}{4}(q-1)-1}-x^{q+1}\cdot x^{-\frac{q-1}{2}}}{x+x^{\frac{3}{4}(q-1)+2}-x^{\frac{q+1}{2}}}
=\frac{1+x^{\frac{1}{4}(q-1)+1}-x^{\frac{q+1}{2}+1}}{x+x^{\frac{3}{4}(q-1)+2}-x^{\frac{q+1}{2}}}\nonumber \\
&=&\frac{(1+x^{\frac{1}{4}(q-1)+1}-x^{\frac{q+1}{2}+1})\cdot x^{\frac{q+1}{2}}}{(x+x^{\frac{3}{4}(q-1)+2}-x^{\frac{q+1}{2}})\cdot x^{\frac{q+1}{2}}}
= -\frac{x^{\frac{q+1}{2}}(x^{\frac{q+3}{4}}+2)^2}{(x^{\frac{q+3}{4}}-2)^2}\\
&=& -\frac{x^{-\frac{q+1}{2}}(x^{\frac{q+3}{4}}+2)^2}{(x^{\frac{q+3}{4}}-2)^2}=\frac{1}{g_1(x)}.
\end{eqnarray*}
Then the conclusion follows from Lemma \ref{q+3div4lem}. \done


\begin{lemma} \label{q-1div2q+3div2lem}
Let $q=5^k$, where $k$ is odd.
 Let $f(x)=-x(\frac{x-2}{x+2})^2 $ and $g(x)=-x(\frac{x+2}{x-2})^2$. Then $f(x)$ permutes $\Omega_{+}$, $g(x) $ permutes $\Omega_{-}$.
\end{lemma}
{\em Proof:}
We will show that $f(x)$ permutes $\Omega_{+}$. Then $g(x) $ permutes $\Omega_{-}$ can be proved in a similar way.

Obviously, if $x\in\Omega_{+}$,  then $(-x(\frac{x-2}{x+2})^2)^{\frac{q+1}{2}}
=-(\frac{x-2}{x+2})^{q+1}=-(\frac{x-2}{x+2})^{q}\cdot \frac{x-2}{x+2}
=-\frac{x^{-1}-2}{x^{-1}+2}\cdot \frac{x-2}{x+2}=1$, which shows that
 $ -x(\frac{x-2}{x+2})^2\in \Omega_{+}$, i.e., $  \{-x(\frac{x-2}{x+2})^2~|~ x\in \Omega_{+} \} \subseteq  \Omega_{+}$.
Suppose that there exist two different  elements $x_1=y_1^2,x_2=y_2^2\in\Omega_{+}$ such that
$ -x_1(\frac{x_1-2}{x_1+2})^2= -x_2(\frac{x_2-2}{x_2+2})^2$.
Then
$y_1(\frac{y_1^2-2}{y_1^2+2})= \pm y_2(\frac{y_2^2-2}{y_2^2+2})$. Let $y_2=uy_1$ for some $u\in \mu_{q+1}$.
\begin{enumerate}
  \item [1)]
  If $y_1(\frac{y_1^2-2}{y_1^2+2})=  y_2(\frac{y_2^2-2}{y_2^2+2})$, then
  $(-y_1^5-2y_1^3)u^3+(y_1^5-2y_1^3)u^2+(2y_1^3-y_1)u+2y_1^3+y_1=0$, i.e.,
$$-y_1(u-1)((y_1^4+2y_1^2)u^2-y_1^2u+2y_1^2+1)=0. $$
Since $y_1\neq 0$, then $u=1$ or $(y_1^4+2y_1^2)u^2-y_1^2u+2y_1^2+1=0$.
Suppose that  $(y_1^4+2y_1^2)u^2-y_1^2u+2y_1^2+1=0$.
Notice that the discriminant of $(y_1^4+2y_1^2)u^2-y_1^2u+2y_1^2+1$ is
$\Delta=2y_1^2(y_1^2-1)^2$, then
$$
u=\frac{y_1^2\pm \sqrt{2}y_1(y_1^2-1)}{2y_1^2(y_1^2+2)}.
$$
Since $q=5^k$ and $k$ is odd, we have $(\sqrt{2})^{q-1}=2^{\frac{q-1}{2}}=(-1)^{\frac{5^k-1}{5-1}}=-1.$
Thus $\sqrt{2}^q=-\sqrt{2}$.
By $u^{q+1}=(\frac{y_1^2\pm \sqrt{2}y_1(y_1^2-1)}{2y_1^2(y_1^2+2)})^{q+1}=1$, we have
\begin{eqnarray*}
(\frac{y_1^2\pm \sqrt{2}y_1(y_1^2-1)}{2y_1^2(y_1^2+2)})^{q+1}&=&(\frac{y_1^2\pm \sqrt{2}y_1(y_1^2-1)}{2y_1^2(y_1^2+2)})^{q}(\frac{y_1^2\pm \sqrt{2}y_1(y_1^2-1)}{2y_1^2(y_1^2+2)})\\
&=&(\frac{y_1^{-2}\mp \sqrt{2}y_1^{-1}(y_1^{-2}-1)}{2y_1^{-2}(y_1^{-2}+2)})(\frac{y_1^2\pm \sqrt{2}y_1(y_1^2-1)}{2y_1^2(y_1^2+2)})=1,
\end{eqnarray*}
then it follows that
\begin{eqnarray}\label{eqq-1q+3eq2}
(y_1^2-1)^2=\pm 2\sqrt{2}y_1(y_1^2-1).
\end{eqnarray}
If $y_1^2=1$, then $u=\frac{y_1^2\pm \sqrt{2}y_1(y_1^2-1)}{2y_1^2(y_1^2+2)}=1$.  Otherwise, take the 2-th power on both sides
of \eqref{eqq-1q+3eq2}, we have
$ y_1^4=-1$, i.e., $x_1=y_1^2=\pm 2$, which is a contradiction since $\pm 2\notin\mu_{q+1} $.
As a result, $u=1$, i.e., $x_1=x_2$.
  \item [2)]
  If $y_1(\frac{y_1^2-2}{y_1^2+2})= - y_2(\frac{y_2^2-2}{y_2^2+2})$, then
  $(y_1^5+2y_1^3)u^3+(y_1^5+3y_1^3)u^2+(y_1+3y_1^3)u+2y_1^3+y_1=0$, i.e.,
$$y_1(u+1)((y_1^4+2y_1^2)u^2+y_1^2u+2y_1^2+1)=0. $$
Since $y_1\neq 0$, then $u=-1$ or $(y_1^4+2y_1^2)u^2+y_1^2u+2y_1^2+1=0$.
Suppose that $(y_1^4+2y_1^2)u^2+y_1^2u+2y_1^2+1=0$.
Notice that the discriminant of $(y_1^4+2y_1^2)u^2+y_1^2u+2y_1^2+1$ is
$\Delta=2y_1^2(y_1^2-1)^2$, then
$$
u=\frac{-y_1^2\pm \sqrt{2}y_1(y_1^2-1)}{2y_1^2(y_1^2+2)}.
$$
Similar as in Case 1), we have
 $$(y_1^2-1)^2=\mp 2\sqrt{2}y_1(y_1^2-1).$$
If $y_1^2=1$, then $u=-1$.  Otherwise, take the 2-th power on both sides  of $y_1^2-1=\mp 2\sqrt{2}y_1$, we have
$ y_1^4=-1$, i.e., $x_1=y_1^2=\pm 2$, which is a contradiction since $\pm 2\notin\mu_{q+1} $.
As a result, $u=-1$, i.e., $y_1=-y_2$ and  $x_1=x_2$.
\end{enumerate}
Based on the above analysis,  we conclude that
 $-x(\frac{x-2}{x+2})^2 $ permutes $\Omega_{+}$. \done

\begin{theorem}\label{th2}
Let $q=5^k$, where $k$ is odd.  Let
$s=\frac{q-1}{2}$  and  $t=\frac{q+3}{2} $.
Then
  $x+x^{s(q-1)+1}-x^{t(q-1)+1} $ permutes $\F_{q^2}$.
\end{theorem}
{\em Proof:}
According to Lemma \ref{lemkey},
it is sufficient to prove that
$g_3(x)=x(1+x^{\frac{q-1}{2}}-x^{\frac{q+3}{2}})^{q-1}$ permutes $\mu_{q+1}$.
If $ 1+x^{\frac{q-1}{2}}- x^{\frac{q+3}{2}} =0$ for some  $x\in \mu_{q+1}$,
 then $ x+x^{\frac{q+1}{2}}- x\cdot x^{\frac{q+3}{2}}=0$. By
$x^{\frac{q+1}{2}}=\pm 1,$ we have
$x+x^{\frac{q+1}{2}}- x\cdot x^{\frac{q+3}{2}}=x\pm 1 \mp x^2=0 $, which
means that $x=\mp 2$, a contradiction with $x\in \mu_{q+1}$ since $\pm 2\notin\mu_{q+1}. $
Thus,   $1+x^{\frac{q-1}{2}}-x^{\frac{q+3}{2}} \neq 0 $ for any $x\in \mu_{q+1}$.
Then we have
\begin{eqnarray*}
g_3(x)=x(1+x^\frac{q-1}{2}-x^{\frac{q+3}{2}})^{q-1}&= &x\frac{(1+x^{\frac{q-1}{2}}-x^{\frac{q+3}{2}})^q}{1+x^{\frac{q-1}{2}}-x^{\frac{q+3}{2}}}\nonumber\\
&=&\frac{x(1+x^{-\frac{q-1}{2}}-x^{-\frac{q+3}{2}})}{1+x^{\frac{q-1}{2}}-x^{\frac{q+3}{2}}}\nonumber \\
&=&\frac{x^2(x^{-1}+x^{-\frac{q+1}{2}}-x^{-2}x^{-\frac{q+1}{2}})}{1+x^{-1}x^{\frac{q+1}{2}}-x\cdot x^{\frac{q+1}{2}}}.
\end{eqnarray*}

If $x\in \Omega_{+}$,  i.e.,
$x^{\frac{q+1}{2}}=1$, then  $g_3(x)=x(1+x^\frac{q-1}{2}-x^{\frac{q+3}{2}})^{q-1}=
\frac{x^2(x^{-1}+x^{-\frac{q+1}{2}}-x^{-2}x^{-\frac{q+1}{2}})}{1+x^{-1}x^{\frac{q+1}{2}}-x\cdot x^{\frac{q+1}{2}}}=
\frac{x^2(x^{-1}+1-x^{-2})}{1+x^{-1}-x}=
-x(\frac{x-2}{x+2})^2 $. By Lemma \ref{q-1div2q+3div2lem}, $g_3(x)$ permutes $\Omega_{+}$.

If
 $x\in \Omega_{-}$,  i.e.,
$x^{\frac{q+1}{2}}=-1$, then  $g_3(x)=x(1+x^\frac{q-1}{2}-x^{\frac{q+3}{2}})^{q-1}=
\frac{x^2(x^{-1}+x^{-\frac{q+1}{2}}-x^{-2}x^{-\frac{q+1}{2}})}{1+x^{-1}x^{\frac{q+1}{2}}-x\cdot x^{\frac{q+1}{2}}}=
\frac{x^2(x^{-1}-1+x^{-2})}{1-x^{-1}+x}=
-x(\frac{x+2}{x-2})^2 $. Again  by Lemma \ref{q-1div2q+3div2lem},  $g_3(x)$  permutes $\Omega_{-}$.
 Then by
  $\Omega_{+} \bigcup \Omega_{-}=  \mu_{q+1} $
and $\Omega_{+} \bigcap \Omega_{-}= \emptyset$,  we have
$g_3(x)=x(1+x^\frac{q-1}{2}-x^{\frac{q+3}{2}})^{q-1}$ permutes $\mu_{q+1}. $
This completes the proof.  \done

\begin{corollary}
Let $q=5^k$, where $k$ is an odd integer.   Then $f(x)=x^q+x^{\frac{q+5}{2}(q-1)+1}-x^{\frac{q+1}{2}(q-1)+1}$
is a permutation trinomial over $\F_{q^2}$.
\end{corollary}
{\em Proof:} According to Lemma \ref{lemkey},
it is sufficient to prove that $g_4(x)=x(x+x^{\frac{q+5}{2}}-x^{\frac{q+1}{2}})^{q-1}$
permutes $\mu_{q+1}$.
Note that
\begin{eqnarray*}
g_4(x)&=&x(x+x^{\frac{q+5}{2}}-x^{\frac{q+1}{2}})^{q-1}=x\frac{x^{-1}+x^{-\frac{q+1}{2}-2}-x^{-\frac{q+1}{2}}}{x+x^{\frac{q+1}{2}+2}-x^{\frac{q+1}{2}}}\nonumber \\
&=&x\frac{x^{-1}+x^{\frac{q+1}{2}-2}-x^{\frac{q+1}{2}}}{x+x^{-\frac{q+1}{2}+2}-x^{-\frac{q+1}{2}}}
=x\frac{x^{-1}+x^{\frac{q+1}{2}-2}-x^{\frac{q+1}{2}}}{x(1+x^{-\frac{q+1}{2}+1}-x^{-\frac{q+3}{2}})}\nonumber \\
&=&\frac{1+x^{\frac{q+1}{2}-1}-x^{\frac{q+3}{2}}}{x(1+x^{-\frac{q+1}{2}+1}-x^{-\frac{q+3}{2}})}=\frac{1}{g_3(x)},
\end{eqnarray*}
where the third equal sign holds due to $x^{-\frac{q+1}{2}}=x^{\frac{q+1}{2}}$.
By Theorem \ref{th2},  $g_3(x)$ permutes $\mu_{q+1}$ if $k$ is odd. This completes the proof.    \done

%

\begin{theorem}\label{th3}
Let $q=5^k$, where $k$ is odd.
Then the following trinomials permute $\F_{q^2}$:
  \begin{enumerate}
    \item [1)]  $x-x^{\frac{(q+3)(q-1)}{2}+1}+x^{q(q-1)+1} $, and
    \item [2)]  $x^q-x^{\frac{q+1}{2}(q-1)+1}+x^{2(q-1)+1}$.
  \end{enumerate}
\end{theorem}

{\em Proof:}
1) According to Lemma \ref{lemkey}, we only need to show that
$g_5(x)= x(1-x^{\frac{q+3}{2}}+x^{-1})^{q-1}$ permutes $\mu_{q+1}$.

Suppose that $x\in \Omega_{+}$, i.e., $x^{\frac{q+1}{2}}=1$, then
$ g_5(x)=x(1-x^{\frac{q+3}{2}}+x^{-1})^{q-1}=x(1-x+x^{-1})^{q-1}=x\cdot \frac{1-x^{-1}+x}{1-x+x^{-1}}$
due to $1-x+x^{-1}\neq 0$ for $x\in \Omega_{+}$. Otherwise, we have   $x(1-x+x^{-1})=-(x^2-x-1)=-(x+2)^2=0$, i.e.,
$x=-2$, a contradiction with $-2\notin \Omega_{+}$. Therefore $g_5(x)=x\cdot \frac{1-x^{-1}+x}{1-x+x^{-1}}
=-x\cdot \frac{x^2+x-1}{x^2-x-1}=-x(\frac{x-2}{x+2})^2$. By Lemma \ref{q-1div2q+3div2lem}, $g_5(x)$  permutes $\Omega_{+}$.

Suppose that $x\in \Omega_{-}$, i.e., $x^{\frac{q+1}{2}}=-1$, then
$ g_5(x)=x(1-x^{\frac{q+3}{2}}+x^{-1})^{q-1}=x(1+x+x^{-1})^{q-1}=x\cdot \frac{1+x^{-1}+x}{1+x+x^{-1}}=x$
due to $1+x+x^{-1}\neq 0$ for $x\in \Omega_{-}$. Otherwise, we have   $x(1+x+x^{-1})=x^2+x+1=0$,
then $(x-1)(x^2+x+1)=x^3-1=0$ for some $x=-y^2\in \Omega_{-}$,
where $y\in \mu_{q+1}$. Thus,  $y^6=-1$ and $y^{12}=1$, together with  $y^{q+1}=1$ and
$\gcd(q+1,12)~|~6$,  we have $y^6=1$, a contradiction.
Therefore, $g_5(x)=x$ permutes $\Omega_{-}$.

Based on the above analysis, we conclude that
 $g_5(x)= x(1-x^{\frac{q+3}{2}}+x^{-1})^{q-1}=x\cdot \frac{1-x^{-\frac{q+3}{2}}+x}{1-x^{\frac{q+3}{2}}+x^{-1}}$ permutes $\mu_{q+1}$.

2) By Lemma \ref{lemkey}, we only need to show that $g_6(x)=x(x-x^{\frac{q+1}{2}}+x^2)^{q-1}$ permutes $\mu_{q+1}$.
Since $x^{\frac{q+1}{2}}=x^{-\frac{q+1}{2}} $, we have
\begin{eqnarray*}
g_6(x)&=&x\cdot\frac{x^{-1}-x^{-\frac{q+1}{2}}+x^{-2}}{x-x^{\frac{q+1}{2}}+x^2}=
x\cdot\frac{x^{-1}-x^{\frac{q+1}{2}}+x^{-2}}{x-x^{-\frac{q+1}{2}}+x^2}\\
&=&
\frac{1-x^{\frac{q+3}{2}}+x^{-1}}{x(1-x^{-\frac{q+3}{2}}+x)}=\frac{1}{g_5(x)}
\end{eqnarray*}
Therefore, $g_6(x) $ permutes $\mu_{q+1}$. \done


Similar as the proof of Theorem \ref{th3}, we have

\begin{theorem}\label{th4}
Let $q=5^k$, where $k$ is odd.
Then the following two trinomials permute $\F_{q^2}$:
  \begin{enumerate}
    \item [1)]  $x-x^{s(q-1)+1}+x^{t(q-1)+1}$, where $s=\frac{q+3}{2}$ and $t= \frac{q+5}{2}$,  and
    \item [2)]  $x^q-x^{s(q-1)+1}+x^{t(q-1)+1},$ where $s=\frac{q+1}{2}$ and $t=\frac{q-1}{2}$.
  \end{enumerate}
\end{theorem}

\begin{lemma}\label{1q-1div2lem}
Let $q=5^k$, where $k$ is even.
 Let  $f(x)= -x(\frac{x-2}{x+2})^2$ and
  $g(x)= -x(\frac{x+2}{x-2})^2$.  Then $f(x)$  permutes $ \Omega_{-}$ and
   $g(x)$  permutes $ \Omega_{+}$.
\end{lemma}
{\em Proof:}
In the following we will prove that  $g(x)$  permutes $ \Omega_{+}$.   Then   $f(x)$  permutes $ \Omega_{-}$
can be proved similarly.
Obviously, if $x\in\Omega_{+}$,  then $(-x(\frac{x+2}{x-2})^2)^{\frac{q+1}{2}}
=-(\frac{x+2}{x-2})^{q+1}=-(\frac{x+2}{x-2})^{q}\cdot \frac{x+2}{x-2}
=-\frac{x^{-1}+2}{x^{-1}-2}\cdot \frac{x+2}{x-2}=1$, which shows that
 $ -x(\frac{x+2}{x-2})^2\in \Omega_{+}$, i.e., $  \{-x(\frac{x+2}{x-2})^2~|~ x\in \Omega_{+} \} \subseteq  \Omega_{+}$.

Suppose that there exist  two different  elements $x_1=y_1^2,x_2=y_2^2\in\Omega_{+}$ such that
$ -x_1(\frac{x_1+2}{x_1-2})^2= -x_2(\frac{x_2+2}{x_2-2})^2$.
Then
$y_1(\frac{y_1^2+2}{y_1^2-2})= \pm y_2(\frac{y_2^2+2}{y_2^2-2})$. Let $y_2=uy_1$ for some $u\in \mu_{q+1}$.
\begin{enumerate}
  \item [1)]
  If $y_1(\frac{y_1^2+2}{y_1^2-2})=  y_2(\frac{y_2^2+2}{y_2^2-2})$, then
  $(4y_1^5+2y_1^3)u^3+(y_1^5+2y_1^3)u^2+(3y_1^3+4y_1)u+3y_1^3+y_1=0$, i.e.,
$$-y_1(u-1)((y_1^4-2y_1^2)u^2+y_1^2u-2y_1^2+1)=0. $$
Since $y_1\neq 0$, then $u=1$ or $(y_1^4-2y_1^2)u^2+y_1^2u-2y_1^2+1=0$.
Suppose that $(y_1^4-2y_1^2)u^2+y_1^2u-2y_1^2+1=0$.
Notice that the discriminant of $(y_1^4-2y_1^2)u^2+y_1^2u-2y_1^2+1$ is
$\Delta=-2y_1^2(y_1^2+1)^2$, then
$$
u=\frac{-y_1^2\pm 2\sqrt{2}y_1(y_1^2+1)}{2y_1^2(y_1^2-2)}.
$$
Since $q=5^k$ and $k$ is even, we have $(\sqrt{2})^{q-1}=2^{\frac{q-1}{2}}=(-1)^{\frac{5^k-1}{5-1}}=1.$
Thus $\sqrt{2}^q=\sqrt{2}$.
By $u^{q+1}=(\frac{-y_1^2\pm 2\sqrt{2}y_1(y_1^2+1)}{2y_1^2(y_1^2-2)})^{q+1}=1$, we have
\begin{eqnarray*}
(\frac{-y_1^2\pm 2\sqrt{2}y_1(y_1^2+1)}{2y_1^2(y_1^2-2)})^{q+1}&=&(\frac{-y_1^2\pm 2\sqrt{2}y_1(y_1^2+1)}{2y_1^2(y_1^2-2)})^{q}(\frac{-y_1^2\pm 2\sqrt{2}y_1(y_1^2+1)}{2y_1^2(y_1^2-2)})\\
&=&(\frac{-y_1^{-2}\pm 2\sqrt{2}y_1^{-1}(y_1^{-2}+1)}{2y_1^{-2}(y_1^{-2}-2)})(\frac{-y_1^2\pm 2\sqrt{2}y_1(y_1^2+1)}{2y_1^2(y_1^2-2)})=1
\end{eqnarray*}
then it follows that
\begin{eqnarray}\label{eq1q-1eq2}
(y_1^2+1)^2=\mp  \sqrt{2}y_1(y_1^2+1).
\end{eqnarray}
If $y_1^2=-1$, then $u=\frac{-y_1^2\pm 2\sqrt{2}y_1(y_1^2+1)}{2y_1^2(y_1^2-2)}=1$.  Otherwise, take the 2-th power on both sides
of \eqref{eq1q-1eq2}, we have
$ y_1^4=-1$, i.e., $x_1=y_1^2=\pm 2$, which is a contradiction since $\pm 2\notin\mu_{q+1} $.
As a result, $u=1$, i.e., $x_1=x_2$.
  \item [2)]
  If $y_1(\frac{y_1^2+2}{y_1^2-2})= - y_2(\frac{y_2^2+2}{y_2^2-2})$, then
  $(y_1^5+3y_1^3)u^3+(y_1^5+2y_1^3)u^2+(2y_1^3+y_1)u+3y_1^3+y_1=0$, i.e.,
$$y_1(u+1)((y_1^4-2y_1^2)u^2-y_1^2u-2y_1^2+1)=0. $$
Since $y_1\neq 0$, then $u=-1$ or $(y_1^4-2y_1^2)u^2-y_1^2u-2y_1^2+1=0$.
Suppose that $(y_1^4-2y_1^2)u^2-y_1^2u-2y_1^2+1=0$.
Notice that the discriminant of $(y_1^4-2y_1^2)u^2-y_1^2u-2y_1^2+1$ is
$\Delta=-2y_1^2(y_1^2+1)^2$, then
$$
u=\frac{y_1^2\pm 2\sqrt{2}y_1(y_1^2+1)}{2y_1^2(y_1^2-2)}.
$$
Similar as in Case 1), we have
 $$(y_1^2+1)^2=\pm  \sqrt{2}y_1(y_1^2+1).$$
If $y_1^2=-1$, then $u=\frac{y_1^2\pm 2\sqrt{2}y_1(y_1^2+1)}{2y_1^2(y_1^2-2)}=-1$.  Otherwise, take the 2-th power on both sides
of $(y_1^2+1)=\pm  \sqrt{2}y_1 $, we have
$ y_1^4=-1$, i.e., $x_1=y_1^2=\pm 2$, which is a contradiction since $\pm 2\notin\mu_{q+1} $.
As a result, $u=-1$, i.e., $ y_1=-y_2$ and  $x_1=x_2$.
\end{enumerate}
Therefore, $-x(\frac{x+2}{x-2})^2 $ permutes $\Omega_{+}$. \done

\begin{theorem}\label{th5}
Let $q=5^k$, where $k$ is even.
Then the following two trinomials permute  $\F_{q^2}$:
  \begin{enumerate}
    \item [1)]  $x-x^{2(q-1)+1}+x^{\frac{(q+3)(q-1)}{2}+1} $, and
    \item [2)]  $x^q-x^{q^2-q+1}+x^{\frac{q+1}{2}(q-1)+1}$.
  \end{enumerate}
\end{theorem}

{\em Proof:}
1) According to Lemma \ref{lemkey}, we only need to show that
$g_7(x)= x(1-x^{2}+x^{\frac{q+3}{2}})^{q-1}$ permutes $\mu_{q+1}$.
First we show that $ 1-x^{2}+x^{\frac{q+3}{2}}=1-x^{2}\pm x \neq 0 $ for any
$x\in\mu_{q+1}$.  Otherwise, $ 1-x^{2}\pm x =-(x\pm 2)^2=0$, i.e., $x=\mp 2$, a contradiction with
$\pm 2\notin \mu_{q+1}$.

Suppose that $x\in \Omega_{+}$, i.e., $x^{\frac{q+1}{2}}=1$, then
\begin{eqnarray*}
 g_7(x)&=&x(1-x^{2}+x^{\frac{q+3}{2}})^{q-1}=x(1-x^{2}+x)^{q-1}\\
 &=&x\cdot \frac{1-x^{-2}+x^{-1}}{1-x^{2}+x}=
-x^{-1}\cdot \frac{x^2+x-1}{x^{2}-x-1}
=-x^{-1}(\frac{x-2}{x+2})^2,
\end{eqnarray*}
thus   $g_7(x)$  permutes $\Omega_{+}$ by Lemma \ref{1q-1div2lem}.

Suppose that $x\in \Omega_{-}$, i.e., $x^{\frac{q+1}{2}}=-1$, then
\begin{eqnarray*}
g_7(x)&=&x(1-x^{2}+x^{\frac{q+3}{2}})^{q-1}=x(1-x^{2}-x)^{q-1}\\
&=&x\cdot \frac{1-x^{-2}-x^{-1}}{1-x^{2}-x}=-x^{-1}\cdot \frac{x^2-x-1}{x^{2}+x-1}
=-x^{-1}(\frac{x+2}{x-2})^2.
\end{eqnarray*}
Again  by Lemma \ref{1q-1div2lem}, $g_7(x)$  permutes $\Omega_{-}$.

Based on the above analysis, we conclude that
 $g_7(x)= x(1-x^{2}+x^{\frac{q+3}{2}})^{q-1}=x\cdot \frac{1-x^{-2}+x^{-\frac{q+3}{2}}}{1-x^{2}+x^{\frac{q+3}{2}}}$ permutes $\mu_{q+1}$.

2) By Lemma \ref{lemkey}, we only need to show that $g_8(x)=x(x-x^{-1}+x^{\frac{q+1}{2}})^{q-1}$ permutes $\mu_{q+1}$.
Since $x^{\frac{q+1}{2}}=x^{-\frac{q+1}{2}} $, we have
\begin{eqnarray*}
g_8(x)&=&x\cdot\frac{x^{-1}-x+x^{\frac{q+1}{2}}}{x-x^{-1}+x^{\frac{q+1}{2}}}=
\frac{1-x^2+x^{\frac{q+3}{2}}}{x-x^{-1}+x^{\frac{q+1}{2}}}\\
&=&
\frac{1-x^2+x^{\frac{q+3}{2}}}{x(1-x^{-2}+x^{\frac{q+1}{2}-1})}=
\frac{1-x^2+x^{\frac{q+3}{2}}}{x(1-x^{-2}+x^{-\frac{q+1}{2}-1})}=\frac{1}{g_7(x)}
\end{eqnarray*}
Therefore, $g_8(x) $ permutes $\mu_{q+1}$.
\done
%

\begin{theorem}\label{th6}
Let $q=5^k$, where $k$ is even.  Let
$s=1$  and  $t=\frac{q-1}{2} $.
Then
  $x+x^{s(q-1)+1}-x^{t(q-1)+1} $ permutes $\F_{q^2}$.
\end{theorem}
{\em Proof:}
According to Lemma \ref{lemkey},
it is sufficient to prove that
$g_9(x)=x(1+x^s-x^{t})^{q-1}$ permutes $\mu_{q+1}$.
First we show that $1+x^s-x^{t}\neq 0$ for any $x\in \mu_{q+1}$.

Suppose that  $1+x^s-x^{t}= 1+x- x^{\frac{q-1}{2}} =0$ for some  $x\in \mu_{q+1}$,
 then $ x+x^{2}- x^{\frac{q+1}{2}}=0$. Note that
$x^{\frac{q+1}{2}}=\pm 1. $ Thus $ x+x^{2}\pm 1 =0$.
If $x+x^{2}- 1=(x-2)^2=0 $, then $x= 2$, a contradiction with $x\in \mu_{q+1}$ since $ 2\notin\mu_{q+1}. $
If $x+x^{2}+1=0 $,
 then $x^3= 1$. Together with  $x^{q+1}=1$ and $\gcd(3,q+1)=1$ due to   $k $ is even, we have   $x=1$,
then we have  $0=x+x^{2}+ 1=3 $, a contradiction.
Therefore,   $1+x-x^{\frac{q-1}{2}} \neq 0 $ for any $x\in \mu_{q+1}$.

Then we have
\begin{eqnarray*}
g_9(x)=x(1+x-x^{\frac{q-1}{2}})^{q-1}&= &x\frac{(1+x-x^{\frac{q-1}{2}})^q}{1+x-x^{\frac{q-1}{2}}}\nonumber\\
&=&\frac{x(1+x^{-1}-x^{-\frac{q-1}{2}})}{1+x-x^{\frac{q-1}{2}}}\nonumber \\
&=&\frac{x^2(x^{-1}+x^{-2}-x^{-\frac{q+1}{2}})}{1+x-x^{-1}\cdot x^{\frac{q+1}{2}}}.
\end{eqnarray*}

If $x\in \Omega_{-}$,  i.e.,
$x^{\frac{q+1}{2}}=-1$, then  $x(1+x-x^{\frac{q-1}{2}})^{q-1}=
\frac{x^2(x^{-1}+x^{-2}-x^{-\frac{q+1}{2}})}{1+x-x^{-1}\cdot x^{\frac{q+1}{2}}}=
\frac{x^2(x^{-1}+x^{-2}+1)}{1+x+x^{-1}}=
x $ permutes $\Omega_{-}$.

If $x\in \Omega_{+}$,  i.e.,
$x^{\frac{q+1}{2}}=1$, then by Lemma \ref{1q-1div2lem},  $x(1+x-x^{\frac{q-1}{2}})^{q-1}=
\frac{x^2(x^{-1}+x^{-2}-x^{-\frac{q+1}{2}})}{1+x-x^{-1}\cdot x^{\frac{q+1}{2}}}=
\frac{x^2(x^{-1}+x^{-2}-1)}{1+x-x^{-1}}=
-x(\frac{x+2}{x-2})^2 $ permutes $\Omega_{+}$.
 Then $g_9(x)=x(1+x-x^{\frac{q-1}{2}})^{q-1}$ permutes $\mu_{q+1}$ due to
  $\Omega_{+} \bigcup \Omega_{-}=  \mu_{q+1} $
and $\Omega_{+} \bigcap \Omega_{-}= \emptyset$.
This completes the proof.  \done
%


Similar as the proof of Theorem \ref{th6}, by using Lemma  \ref{1q-1div2lem},  we can construct
the following  permutation trinomials over $\F_{5^n}$.
\begin{theorem}\label{th7}
Let $q=5^k$, where $k$ is even.
Then the following trinomials  permute  $\F_{q^2}$:
\begin{enumerate}
  \item [1)]  $x-x^q+x^{\frac{q+5}{2}\cdot (q-1)+1} $,
  \item [2)]  $x+x^{\frac{q+3}{2}\cdot (q-1)+1}+x^{\frac{q+5}{2}\cdot (q-1)+1}$, and
  \item [3)]  $x+x^{\frac{q+3}{2}\cdot (q-1)+1}-x^{q^2-q+1}$.
\end{enumerate}
\end{theorem}

It is known that the inverse of a Niho exponent, if exists,  is again a Niho exponent,
and the product of two Niho exponents is still a Niho exponent \cite{Dobbertin99}.
Similar as Lemma 4 in \cite{Li161}, we have the following lemma.

\begin{lemma}\label{lemequal}
Let $q=p^k$, where $p$ is a prime and $k$ is an integer.
 \begin{enumerate}
   \item [1)]   Suppose that  $f(x)=x(1+x^i+x^j)^{q-1}$ permutes
 $\mu_{q+1}$.  If $\gcd(2i-1,q+1)=1$,  then  $x(1+x^s+x^t)^{q-1}$ permutes
 $\mu_{q+1}$,  where $s=\frac{i}{2i-1} $ and $ t=\frac{i-j}{2i-1}$.
   \item  [2)] Suppose that   $f(x)=x(1+x^i-x^j)^{q-1}$ permutes
 $\mu_{q+1}$.  If $\gcd(2i-1,q+1)=1$, then $x(1+x^s-x^t)^{q-1}$ permutes
 $\mu_{q+1}$,  where $s=\frac{i}{2i-1}$ and
 $ t=\frac{i-j}{2i-1}$; if $\gcd(2j-1,q+1)=1$, then $x(1-x^s-x^t)^{q-1}$ permutes $\mu_{q+1}$,  where $s=\frac{j}{2j-1} $
  and $t=\frac{j-i}{2j-1}$.
  \item [3)]   Suppose that  $f(x)=x(1-x^i-x^j)^{q-1}$ permutes
 $\mu_{q+1}$.  If  $\gcd(2i-1,q+1)=1$,   then $x(1-x^s+x^t)^{q-1}$   permutes
 $\mu_{q+1}$,
  where $s=\frac{i}{2i-1}$  and  $t=\frac{i-j}{2i-1}$.
  \end{enumerate}
\end{lemma}
The integers $s$ and $t$ written as fractions in Lemma \ref{lemequal}  should be interpreted as modulo $p^k+1$.
Based on Lemma \ref{lemequal}, one can obtain lots of permutation trinomials over $\F_{5^n}$ from the permutation
trinomials constructed in this section.

In Table \ref{table1},  we list all the permutation trinomials  of the form \eqref{polyform}
constructed in this paper.   We denote by $(\pm [c_1],\pm [c_2])$  the permutation trinomial
$f(x)=x\pm x^{c_1(5^k-1)+1}\pm x^{c_2(5^k-1)+1}.$  For example,
the pair  $(+[1],-[\frac{5^k-1}{2}])$ means the trinomial
$ x+x^{5^k}-x^{\frac{5^k-1}{2}(5^k-1)+1}$. The ``Equivalent Pairs" column in Table \ref{table1} are obtained based on Lemma \ref{lemequal}.

\begin{table}[h!]
  \begin{center}
  \small{
  \caption{Pairs  $(\pm [c_1],\pm [c_2])$ such that $f(x)$ defined by \eqref{polyform} are permutation trinomials over $\F_{q^2}$ ($q=5^k$)}
\label{table1}
    \begin{tabular}{| l| c |c| c|c|}
    \hline
$(\pm [c_1],\pm [c_2])$                 &   $  g(x)$       &Conditions          &   Equivalent Pairs   &  Proven in          \\
  \hline\hline
$(+[\frac{q+3}{4}],-[\frac{q+3}{2}]) $  & $-x^{\frac{q+1}{2}}(\frac{x^s-2}{x^s+2})^{2}$ ($s=\frac{q+3}{4}$)    &  for all $k$  &$(-[\frac{q+3}{4}],-[\frac{q+3}{2}])$  & Theorem \ref{th1}   \\
   \hline
$(+[\frac{q-1}{2}],-[\frac{q+3}{2}])$  &$\frac{-x(x^{\frac{q+3}{2}}-x^{\frac{q-1}{2}}+1)}{x^{\frac{q+3}{2}}-x^{\frac{q-1}{2}}-1}$&    $k $ odd          &  $(-[2],-[\frac{q+3}{2}])$ & Theorem \ref{th2}    \\
  \hline
$(+[-1],-[\frac{q+3}{2}])$ &$\frac{x^2(x^{\frac{q-1}{2}}-x-1)}{x^{\frac{q+5}{2}}-x-1}$&        $k $ odd     &$(-[\frac{q+3}{2}],-[\frac{q+5}{2}])  $  &  Theorem \ref{th3}     \\
 \hline
$(-[\frac{q+3}{2}],+[\frac{q+5}{2}])  $&$\frac{-x(x^{\frac{q-1}{2}}-x^{\frac{q-3}{2}}-1)}{x^{\frac{q+5}{2}}-x^{\frac{q+3}{2}}+1}$& $k $ odd   &  $(-[-1],-[\frac{q+3}{2}])$ &  Theorem \ref{th4}    \\
 \hline
$(-[2],+[\frac{q+3}{2}])$  &$\frac{x^{\frac{q+3}{2}}+x^2-1}{x(x^{\frac{q+3}{2}}-x^2+1)}$ &   $k $ even    & $(+[\frac{q+3}{2}],-[\frac{q-1}{2}])  $, & Theorem \ref{th5}  \\
  & &   &  $(-[2\cdot \frac{q+2}{3}],-[\frac{q+2}{3}\cdot\frac{q+3}{2}])$ &   \\
 \hline
$(+[1],-[\frac{q-1}{2}])$  &$\frac{x^{\frac{q+5}{2}}-x-1}{x^{\frac{q-1}{2}}-x-1}$&   $k $ even    & $(+[1],-[\frac{q+5}{2}])  $, & Theorem \ref{th6}  \\
  & &   &  $(-[\frac{q+2}{3}\cdot\frac{q+5}{2}],-[\frac{q+2}{3}\cdot\frac{q+3}{2}])$ &   \\
 \hline
 $(-[1],+[\frac{q+5}{2}])$  &$\frac{x^{\frac{q-1}{2}}+x-1}{x^{\frac{q+5}{2}}-x+1}$&   $k $ even    & $(-[1],-[\frac{q-1}{2}])  $, & Theorem \ref{th7}  \\
   & &   &   $(+[\frac{q+2}{3}\cdot\frac{q+5}{2}],-[\frac{q+2}{3}\cdot\frac{q+3}{2}])$ &   \\
 \hline
  $(+[\frac{q+3}{2}],+[\frac{q+5}{2}])$  &$\frac{x^{\frac{q+1}{2}}+x^{\frac{q-1}{2}}+x}{x^{\frac{q+5}{2}}+x^{\frac{q+3}{2}}+1}$&   $k $ even    & $(+[\frac{q+3}{2}],+[-1])  $,& Theorem \ref{th7}  \\
     & &   &    $(+[\frac{q+2}{3}\cdot\frac{q+5}{2}],+[\frac{q+2}{3}])$ &   \\
 \hline
    $(+[\frac{q+3}{2}],-[-1])  $ &$ \frac{x^2(x^{\frac{q-1}{2}}-x+1)}{x^{\frac{q+5}{2}}+x-1}$&   $k $ even    & $(+[\frac{q+3}{2}],-[\frac{q+5}{2}])$, & Theorem \ref{th7}  \\
         & &   &    $(-[\frac{q+2}{3}\cdot\frac{q+5}{2}],-[\frac{q+2}{3}])$ &   \\
 \hline
   $(+[2],-[-2])$  &$-x(\frac{x^2+2}{x^2-2})^2$    &   $k $ odd    & $(-[-2\cdot 5^{k-1}],-[-4\cdot 5^{k-1}])$& Proposition  \ref{pro1}  \\
 \hline
    $(-[\frac{q+5}{3}],-[ \frac{2(q+2)}{3}])$  &$-x(\frac{x^2-2}{x^2+2})^2$ &   $k $ even     & $(+[-2\cdot 5^{k-1}],-[-4\cdot 5^{k-1}])$& Proposition  \ref{pro2}  \\
 \hline
\end{tabular}
}
\end{center}
\end{table}

To end this section,  we propose  two  conjectures. If the conjectures are  true, then
we can construct two classes  of permutation trinomials over $\F_{5^n}. $

\begin{conjecture}\label{conj1}
The polynomial  $x(\frac{x^2-x+2}{x^2+x+2})^2$ is a permutation polynomial  over $\F_{5^k}$ for odd $k$.
\end{conjecture}

\begin{remark}
Let $\Lambda_1=\{x^2~|~x \in \F^*_{5^k}\}$ and $\Lambda_2=\{2x^2~|~x \in \F^*_{5^k}\}$, where $k$ is odd.   
 Note that $\F_{5^k}^*=\Lambda_1 \bigcup \Lambda_2$ and $
 \Lambda_1\bigcap \Lambda_2 = \emptyset.$
 Obviously,
$\{ x(\frac{x^2-x+2}{x^2+x+2})^2~|~x \in \Lambda_1  \} \subseteq  \Lambda_1$ and
$\{ x(\frac{x^2-x+2}{x^2+x+2})^2~|~x \in \Lambda_2  \} \subseteq  \Lambda_2$. 
 Thus, to prove Conjecture \ref{conj1} is equivalent to prove that
  $\{ x(\frac{x^2-x+2}{x^2+x+2})^2~|~x \in \Lambda_1  \}\supseteq\Lambda_1$.
\end{remark}

\begin{proposition}\label{pro1}
 Let $n=2k$, where $k$ is an odd integer. Let $q=5^k$.  If Conjecture \ref{conj1} is true, then
$f(x)=x+x^{2(q-1)+1}-x^{(q-1)(q-1)+1}$ permutes $\F_{q^2}$.
\end{proposition}
{\em Proof:}
First it is easy to show that if $x\in\F_{5^n}\setminus \F_{5^k}$, then
 $f(x)=x+x^{2q-1}-x^{3-2q}\in\F_{5^n}\setminus \F_{5^k}$. Otherwise,
 if there exists $x_1 \in\F_{5^n}\setminus \F_{5^k}$ such that $f(x_1)\in\F_q$, then $f(x_1)=f(x_1)^q$, i.e.,
  $$x_1+x_1^{2q-1}-x_1^{3-2q}=x_1^q+x_1^{2-q}-x_1^{3q-2}.$$ Let
 $y=x_1^{q-1}$, we get    $1+y^2-y^{-2}=y+y^{-1}-y^{3}$, thus,
 $$y^5+y^4-y^3+y^2-y-1=(y-1)(y^4 + 2y^3 + y^2 + 2y + 1)=0.$$
 Suppose that $y^4 + 2y^3 + y^2 + 2y + 1= 0$, then the order of $y$ is $13$, thus $13~|~(q+1)$, which
is a contradiction with
 $\gcd(13, q+1)=1$ due to $k$ is odd.
Therefore, we have  $y=1$, i.e., $x_1\in\F_q$, a contradiction.

%
%
%
%
%
%

In the following we show that $f(x)=c$ has at most one solution for any $c\in\F_{q^2}$.

If $ c\in\F_q$, then $x\in\F_q$, thus we have $f(x)=x=c$. Therefore, $x=c$ is the only solution of $f(x)=c$.

If $c\in \F_{q^2}\setminus \F_{q}$, then $x\in \F_{q^2}\setminus \F_{q}$, and it follows that
$\N(x)\neq 0$. We compute
\begin{eqnarray*}
\Tr(f(x))&=&\Tr(x+x^{2q-1}-x^{-2q+3})\nonumber \\
         &=&\Tr(x)+(x^{2q-1}+x^{2-q})-(x^{-2q+3}+x^{-2+3q}) \nonumber \\
         &=&\Tr(x)+\frac{x^{3}+x^{3q}}{x^{q+1}}-\frac{x^{5}+x^{5q} }{x^{2+2q}}\nonumber\\
         &=&\Tr(x)+\frac{\Tr(x^3)}{\N(x)}-\frac{\Tr^5(x)}{\N^2(x)},
\end{eqnarray*}
and
\begin{eqnarray*}
\N(f(x))&=&\N(x+x^{2q-1} -x^{-2q+3})\\
         &=&(x+x^{2q-1} -x^{-2q+3})(x^q+x^{2-q}-x^{-2+3q} )\\
         &=&3x^{1+q}-(x^{5q-3}+x^{5-3q})\\
         &=&3\N(x)-\frac{x^{8q}+x^{8}}{x^{3q+3}}\\
         &=&3\N(x)-\frac{\Tr(x^8)}{\N^3(x)}
\end{eqnarray*}
Denote  $a=\Tr(x)$, $b=\N(x)$,
$\alpha=\Tr(f(x))$,  and  $\beta=\N(f(x))$.  According to Lemma \ref{tracelemma},  we have
\begin{eqnarray}\label{eqalpha2-2}
\alpha=\Tr(f(x))&=&\Tr(x)+\frac{\Tr(x^3)}{\N(x)}-\frac{\Tr^5(x)}{\N^2(x)}\nonumber \\
&=&a+\frac{a^3+2ba}{b}-\frac{a^5}{b^2},\nonumber \\
&=&a(3+\frac{a^2}{b}-\frac{a^4}{b^2}),
\end{eqnarray}
and
\begin{eqnarray}\label{eqbeta2-2}
\beta=\N((x))&=&3\N(x)-\frac{\Tr(x^8)}{\N^3(x)}\nonumber \\
&=&3b-\frac{a^8+2ba^6-b^3a^2+2b^4}{b^3}\nonumber \\
&=&b(1-\frac{a^8}{b^4}-2\cdot \frac{a^6}{b^3}+\frac{a^2}{b})
\end{eqnarray}


Denote  $\gamma=\frac{\alpha^2}{\beta}$, then by \eqref{eqalpha2-2} and \eqref{eqbeta2-2}, we have
$\gamma=-r(\frac{r^2-r+2}{r^2+r+2})^2$, where $r=\frac{a^2}{b}$.  If Conjecture \ref{conj1} is true,
  we know that
 $r$ is uniquely determined by $\gamma$, thus $r$  is uniquely determined by $c$.
  From \eqref{eqalpha2-2} and \eqref{eqbeta2-2}, it follows that $a=\Tr(x)$ and $b=\N(x)$ are uniquely determined by $c$.
This completes the proof since  $(x,x^q)$  is uniquely determined by  $a=\Tr(x)$ and $b=\N(x)$, and $f(x)\neq f(x)^q=f(x^q)$.
 \done

\begin{remark}
By Lemma \ref{lemkey}, for odd $k$,  $f(x)=x+x^{2(q-1)+1}-x^{(q-1)(q-1)+1}$ permutes
$\F_{q^2}$ if and only if $g(x)=x(1+x^2-x^{-2})^{q-1}
=x\cdot \frac{1+x^{-2}-x^{2}}{1+x^2-x^{-2}}
=-x(\frac{x^2+2}{x^2-2})^2$ permutes $\mu_{q+1}$.  Thus, if Conjecture \ref{conj1} is true,
then $-x(\frac{x^2+2}{x^2-2})^2$ permutes $\mu_{q+1}$.
\end{remark}

\begin{conjecture}\label{conj2}
Let $q=5^k$, where $k$ is an even integer.  Then  $-x(\frac{x^2-2}{x^2+2})^2$ permutes $\mu_{q+1}$.
\end{conjecture}

\begin{proposition}\label{pro2}
 Let $n=2k$, where $k$ is an even integer. Let $q=5^k$.  If Conjecture \ref{conj2} is true, then
$f(x)=x-x^{s(q-1)+1}-x^{t(q-1)+1}$ permutes $\F_{q^2}$, where $s=\frac{q+2}{3}+1$ and $t=2\cdot\frac{q+2}{3}$.
\end{proposition}
{\em Proof:}
According to Lemma \ref{lemkey}, we only need to prove that
$g_{10}(x)= x(1-x^{s}-x^{t})^{q-1}$ permutes $\mu_{q+1}$,
which  is equivalent to showing $ g_{10}(x^3)= x^3(1-x^{3s}-x^{3t})^{q-1}=
x^3(1-x^{q+2+3}-x^{2(q+2)})^{q-1}=x^3(1-x^4-x^2)^{q-1}$ permutes $\mu_{q+1}$
   due to $\gcd(3,q+1)=1.$
   Notice that
   $ g_{10}(x^3)=x^3(1-x^4-x^2)^{q-1}=x^3\cdot \frac{1-x^{-2}-x^{-4}}{1-x^2-x^4}=-x^{-1}\frac{x^4-x^2-1}{x^4+x^2-1}=
   -x^{-1}\cdot (\frac{x^2+2}{x^2-2})^2$. Therefore, if Conjecture \ref{conj2} is true,
   $f(x)=x-x^{s(q-1)+1}-x^{t(q-1)+1}$ permutes $\F_{q^2}$. \done


\section{Conclusion} \label{conclu}
In this paper, we find some new fractional polynomials that  permute the  set  of the  $(q+1)$-th roots of
unity in $\F_{q^2}$, where $q=5^k$.
 Based on these fractional polynomials, we construct several classes
permutation trinomials over $\F_{5^{2k}}$.
Two conjectures
 are  proposed, and
 two new classes  of permutation trinomials over $\F_{5^{2k}}$ will be obtained if the two conjectures are  true.
At the end of this paper, we propose two problems about the permutation trinomials with the
form  \eqref{polyform},  it will be nicer if they can be solved.

{\em Problem 1:} Let $s+t=0 $ or $s+t=\frac{5^k+1}{2}$.  Under what conditions for $s$ is the
 polynomial $x+x^{s(5^k-1)+1}-x^{t(5^k-1)+1}$ a permutation
trinomial over $\F_{5^{2k}} $?

{\em Problem 2:} Let $s+t=\frac{5^k+1}{2}$.  Under what conditions for $s$ is the
polynomial $x+x^{s(5^k-1)+1}+x^{t(5^k-1)+1}$ a permutation
trinomial over $\F_{5^{2k}} $?

%
%

\end{document}